\documentclass[pra,reprint,superscriptaddress,twocolumn,showkeys]{revtex4-1}
\usepackage[utf8]{inputenc} 
\usepackage{amsmath}
\usepackage{braket}
\usepackage{graphicx}
\usepackage{subfigure}
\usepackage{pgfplots}
\usepackage{csquotes}
\usepackage{hhline}
\usepackage{amssymb}

\usepackage{dcolumn}
\usepackage{tabularx}
\usepackage[colorlinks=true,linkcolor=blue,citecolor=blue,urlcolor=blue]{hyperref}
\usepackage{longtable}
\usepackage{braket}
\usepackage{float}
\usepackage{listings}
\usepackage{color}
\definecolor{mygreen}{rgb}{0,0.6,0}
\definecolor{mygray}{rgb}{0.5,0.5,0.5}
\definecolor{mymauve}{rgb}{0.58,0,0.82}
\usepackage{amssymb}
\usepackage{epstopdf}
\usepackage{xcolor}

\newcolumntype{C}{>{\centering\arraybackslash}X}
\begin{document}

\title{Single photon generation and non-locality of perfect W-state}

\author{Manoranjan Swain}
\email{swainmanoranjan333@gmail.com}
\author{Amit Rai}
\email{amitrai007@gmail.com}
\affiliation{Department of Physics and Astronomy, National Institute of Technology, Rourkela, 769008, Odisha, India}

\author{M. Karthick Selvan}
\email{karthick.selvan@yahoo.com}
\affiliation{Department of Physics, Thiagarajar College, Madurai-625009, Tamilnadu, India}

\author{Prasanta K. Panigrahi}
\email{pprasanta@iiserkol.ac.in}
\affiliation{Department of Physical Sciences,\\ Indian Institute of Science Education and Research Kolkata, Mohanpur 741246, West Bengal, India}

\begin{abstract}
 We study the generation of single photon perfect W-state. An important aspect of this perfect W-state is that, it can be used for perfect teleportation and superdense coding, which are not achievable with maximally entangled W-state. Our scheme for generation involves entanglement between various path degrees of freedom of a single photon in a compact and weakly coupled integrated waveguide system, which can be fabricated precisely with femtosecond laser direct writing technique. These platforms are interferometrically stable, scalable, less sensitive to decoherence and ensures a very low loss factor of 0.1dBcm$^{-1}$ during photon propagation and hence are ideal for generation of perfect W-state. In addition to generation of single photon perfect W-state we study its non local properties using theory of local elements of reality. 
\end{abstract}

\maketitle

\section{Introduction}

Multiqubit entangled states~\cite{pnf_2,pnfl_4,pnf_7,pnfl_25, pnfl_26}, due to their important applications in quantum information science, are extensively studied these days. One of these states is GHZ state~\cite{pnf_1,pnf_2}, which was used to verify Bell's theorem~\cite{pnf_3} without the involvement of any inequality. Studies on this state have revealed extreme non-classical properties of it~\cite{pnf_2}. Its important applications can be found in Ref.~\cite{pnf_4,pnf_5,pnf_6}. However when losses are considered this state is seen to be fragile in carrying entanglement. There exist one more type of multiqubit entangled state known as W-state~\cite{pnf_7}. This state is known to have robust properties against particle loss and it can carry entanglement for a longer time. These two multiqubit states are shown to be nonequivalent to each other under local operation and classical communication~\cite{pnf_7}. Applications of W-state can be found in Ref.~\cite{pnf_8,pnf_9,pnf_10,pnfl_10}. The general form of the W-state~\cite{pnfl_1} can be given as:

\begin{equation}\label{eq1}
\begin{split}
&\arrowvert W\rangle= \sum C_i |1_i,\{0\}\rangle,\\
\hspace{1cm} &\sum{|C_i|^2}=1.
\end{split}
\end{equation}
where $|1_i,\{0\}\rangle$ represents the qubit state $|1\rangle$ in $i$th mode and $|0\rangle$ in all other modes. With different combinations of $C_i$'s, different forms of W-state can be realized. For example, if one chooses $C_i$=$\frac{1}{\sqrt{P}}$(for P number of waveguide modes), maximally entangled W-state is obtained~\cite{w1,w2}.  However studies on teleportation protocol with W-state suggested non unit fidelity~\cite{pnf_11}. In other words teleportation processes with maximally entangled W-states are not 100$\%$ perfect. In their paper Agrawal and Pati~\cite{pnf_12} have suggested the existence of another kind of non-maximally entangled asymmetric three mode W-state which is suitable for perfect teleportation and superdense coding and is known as perfect W-state. The generalized perfect W-state is given as:
\begin{equation}\label{eq2}
  \arrowvert W_{p,s}\rangle= \frac{1}{\sqrt{2+2s}}(\arrowvert100\rangle+ \sqrt{s}e^{i\Phi_1}\arrowvert010\rangle+ \sqrt{s+1}e^{i\Phi_2}\arrowvert001\rangle)
\end{equation}
where s is a real number, $\Phi_1$ and $\Phi_2$ are phases acquired by the state. Using such asymmetric W-state not only perfect teleportation~\cite{pnf_12} but also splitting, sharing and transmission of quantum information are shown~\cite{pnf_13,pnf_14}. Consequently generation of this state is essential in the frame work of quantum computation and information processing. Li Dong $\emph{et al}$ has proposed a scheme of generation for such perfect W-state using polarization entanglement of three photons in bulk optical elements~\cite{pnf_16}. However such arrangements are large in size, unscalable and are sensitive to decoherence.

 In this paper we present a scheme for the generation of three mode perfect W-state by sharing a single photon between different optical modes. The qubit state is realized by the presence or absence of photon in corresponding mode. The photonic fock state i.e. the vacuum (excited) state represents qubit state $\arrowvert0\rangle$ ($\arrowvert1\rangle$). Earlier work~\cite{pnf_16} has considered the multiparticle perfect W-state. Note that non-locality and entanglement are usually considered as the behaviour of two or more spatially separated particles (qubits in present case). Generally in photonic systems polarization (H or V) is considered as the two levels to represent a qubit state. In such case the number of particles required is exactly same as the number of qubits used to represent a quantum state. However Tan \emph{et al}~\cite{pnf_32} first pointed out the fact that non-locality and entanglement can also be verified for a single particle (photon) superposition over two different spatial field modes. Later this was experimentally demonstrated by Hessmo \emph{et al} ~\cite{pnf_33}. Recently Papp \emph{et al} ~\cite{new1} have investigated the multipartite maximally entangled W-state where single photon is shared among four different optical modes and experimentally demonstrated that these modes are entangled. Moreover Monteiro \emph{et al} ~\cite{new2} have proposed a scalable scheme of detecting entanglement between multiple optical paths sharing a single photon using local measurements and small displacement operations. We would like to further emphasize that, the entanglement containing single photon between its subsystems has found interesting applications such as, deterministic quantum teleportation~\cite{new10}, path entanglement based teleportation~\cite{new3}, entanglement swapping~\cite{new4}, entanglement purification~\cite{new5}, demonstration of Einstein-Podolsky-Rosen Steering~\cite{new11}, quantum repeaters~\cite{new12} etc. Moreover Ref.~\cite{r1,r2,r3,r4} discuss similar concepts in the classical regime.

The platform we have chosen is a system of coupled integrated waveguides. This system is compact, less sensitive towards decoherence and are interferometrically stable. Fabrication of these types of systems have been precisely shown~\cite{pnf_17,pnf_18} by using femtosecond laser direct-writing technique. Furthermore loss in photon propagation for these systems was shown to be very low i.e. about 0.1dB $cm^{-1}$~\cite{pnf_19,pnf_20}. Due to these important properties photonic integrated circuits~\cite{pnf_21,pnf_22,pnf_23,pnf_24,pnf_25,pnf_26,pnf_27,pnf_28,pnf_29,pnf_30} are finding application in quantum computation as well as in quantum information science. Politi \emph{et al} have shown high fidelity optical realizations of quantum photonic circuits and quantum gates using silica-on-silicon integrated systems~\cite{pnf_23}. Control and manipulation of single photon and multi photon entangled states were shown to be possible by introducing phase shifters in integrated interferometers~\cite{pnf_31}. Hence these systems are ideal for generation of perfect W-state. We also analyzed the effect of decoherence in our scheme, which confirms the robustness of the scheme at the presence of loss. Along with generation we also study the nonlocal properties of such single particle perfect W-state.

The description of our work in this paper has been divided into different sections. Section II describes a detailed process of generation of the desired state with relevant discussions on it in Section III. Section IV gives idea about the nonlocality study on the state followed by conclusion in section V.

\section{Generation of perfect W- state}
\begin{figure}[h]
 \subfigure[]{\includegraphics[width=0.95\linewidth]{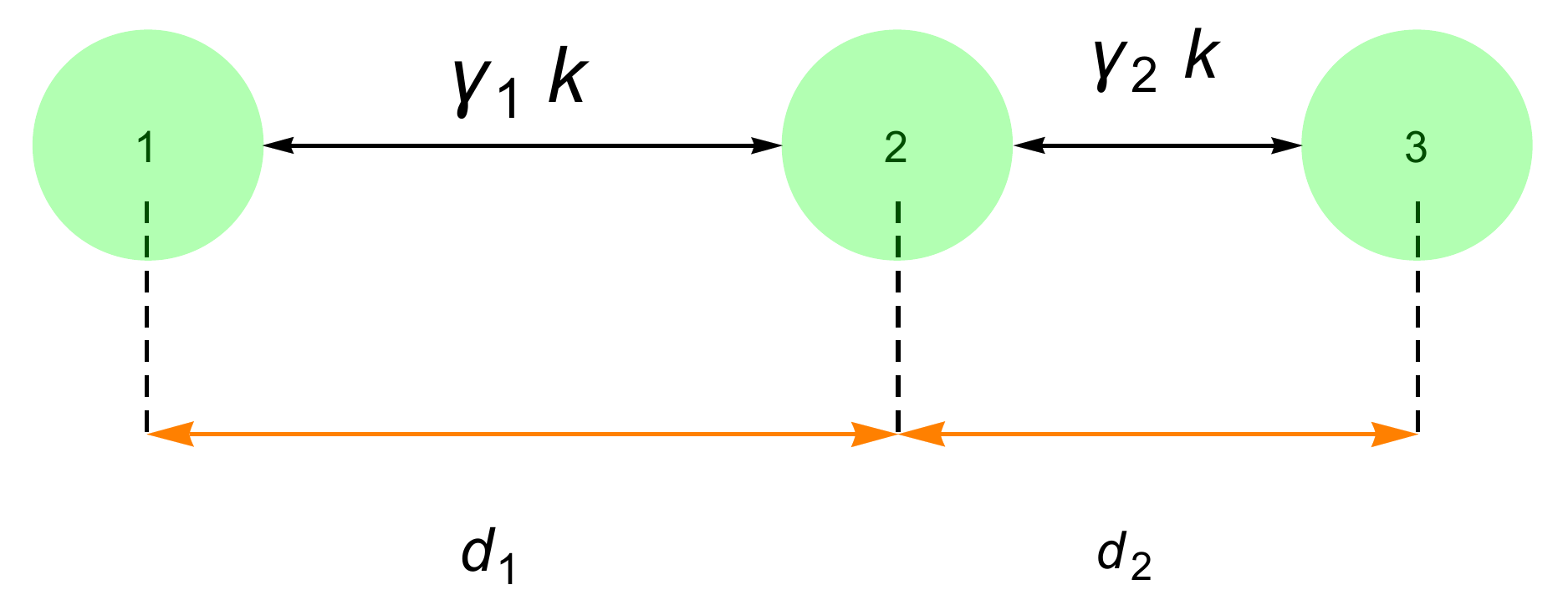}}
  \subfigure[]{\includegraphics[width=0.9\linewidth]{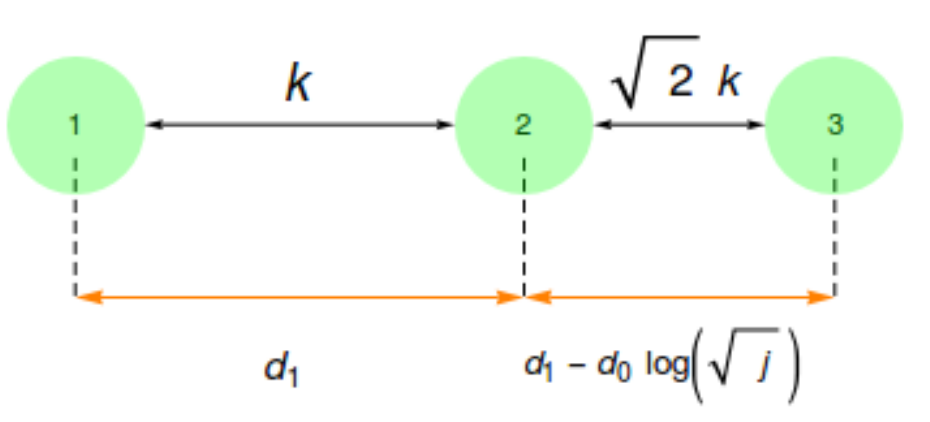}}
  \caption{(a) Schematic view of cross section of coupled waveguide system. Coupling parameters between the guides, $\gamma_1 k$ and $\gamma_2 k$ depends on d$_1$ and d$_2$ respectively. (b) Desired system for generation of $W_p$-state with s=1, $\Phi_1$=0, and $\Phi_2=0$.}
  \label{wav}
\end{figure}

The general form of perfect W-state is represented in Eq. $\eqref{eq2}$. To produce such a state with a single photon we have considered an integrated photonic system. The system consists of three identical single-mode weakly coupled waveguide structures as depicted in Fig. \ref{wav}(a). Note that the coupling parameter between the guides (1, 2) and (2, 3) is kept different. The coupling parameter between adjacent guides depends on the spatial separation and can be adjusted as required, by varying the relative separation between them.

The Hamiltonian of the system can be written as,
\begin{equation}
 \hat{H}=\hbar\omega\sum_{j=1}^N\hat{a}_j^\dagger\hat{a}_j+ \hbar \sum_{j=1}^{N-1}k'(\hat{a}_j^\dagger \hat{a}_{j+1}+\hat{a}_j \hat{a}_{j+1}^\dagger )
\end{equation}
where the first term represents the free propagation of light with $\omega$ proportional to refractive index of the material. $\hat{a}_j^\dagger(\hat{a}_j)$ is the bosonic creation(annihilation) operator. $k'$ represents the coupling coefficient between different guides. The value of $k'$ is set to be $k'=k\gamma_j$, where $k$ is the coupling strength and $\gamma_j$ is a number that depends on j (waveguide index). With this the corresponding Heisenberg equation of motion is given as,

\begin{equation}\label{m}
 i \frac{d\hat{a}_j^\dagger}{dz}= k\gamma_{j-1} \hat{a}_{j-1}^\dagger+ k \gamma_{j}\hat{a}_{j+1}^\dagger
\end{equation}

The Heisenberg equation is expressed as a function of z here, as it is related to propagation time `t' by $t=z\mu/c$ ($\mu$ is the refractive index of the waveguide medium). In matrix form the Heisenberg equation for creation operators is given as:
\begin{equation}
 i\frac{dA^\dagger}{dz}= M A^\dagger
\end{equation}
where A$^\dagger$ is the set of creation operators of each guide and M is the coupling matrix. With the assumption of identical guides, the individual propagation constants are neglected from the equations as they can only introduce a global phase in the result.

The generalized perfect W-state given in Eq. $\eqref{eq2}$ contains a class of asymmetric W-states with different values of s, $\Phi_1$ and $\Phi_2$.
For simplicity we choose s=1, $\Phi_1$=0 and $\Phi_2$=0 and the resultant state can be represented as,
\begin{equation}\label{eq3}
  \arrowvert W_{p,1}\rangle= \frac{1}{2}(\arrowvert100\rangle+ \arrowvert010\rangle+ \sqrt{2}\arrowvert001\rangle)
\end{equation}
To obtain the above mentioned (in Eq. $\eqref{eq3}$) perfect W-state, the value of $\gamma_j$ is chosen to be $\sqrt{j}$. With this the Heisenberg equations of motion (in Eq. $\eqref{m}$) simplifies to,
\begin{equation}
 i \frac{d\hat{a}_j^\dagger}{dz}= k\sqrt{j-1} \hat{a}_{j-1}^\dagger+ k \sqrt{j}\hat{a}_{j+1}^\dagger
\end{equation}
and in matrix form,
\begin{equation}
i\frac{d}{dz}
 \begin{pmatrix}
\vspace{0.2cm} a_1^\dagger \\ \vspace{0.2cm} a_2^\dagger\\a_3^\dagger

 \end{pmatrix}
= 
\begin{pmatrix}
 0 & k & 0\\k & 0 & \sqrt{2} k\\0 & \sqrt{2} k & 0
\end{pmatrix}
\begin{pmatrix}
 \vspace{0.2cm} a_1^\dagger \\ \vspace{0.2cm} a_2^\dagger\\a_3^\dagger

\end{pmatrix}
\end{equation}

As the coupling matrix M is assumed to be independent of z, the input and the output states are related by the evolution matrix as,
\begin{equation}
 A^\dagger(z)=e^{-izM}A^\dagger(0)
\end{equation}

where $e^{-izM}$ is the evolution matrix and can be written as:
\begin{equation}
\scriptstyle e^{-izM}=
\begin{pmatrix}
\scriptstyle\frac{1}{3}(2+\text{cos}[\sqrt{3}kz]) & -\scriptstyle\frac{i \text{sin}[\sqrt{3}kz]}{\sqrt{3}}& \scriptstyle \frac{\sqrt{2}}{3}(-1+\text{cos}[\sqrt{3}kz])\\  -\scriptstyle\frac{i \text{sin}[\sqrt{3}kz]}{\sqrt{3}} & \scriptstyle \text{cos}[\sqrt{3}kz] &  -\scriptstyle  \frac{i \sqrt{2}}{\sqrt{3}}\text{sin}[\sqrt{3}kz] \\ \scriptstyle\frac{\sqrt{2}}{3}(-1+\text{cos}[\sqrt{3}kz]) &  -\scriptstyle \frac{i \sqrt{2}}{\sqrt{3}}\text{sin}[\sqrt{3}kz] & \scriptstyle\frac{1}{3}(1+2\text{cos}[\sqrt{3}kz])
\end{pmatrix}
\end{equation}

\begin{figure}[h]

  \includegraphics[width=0.47\textwidth]{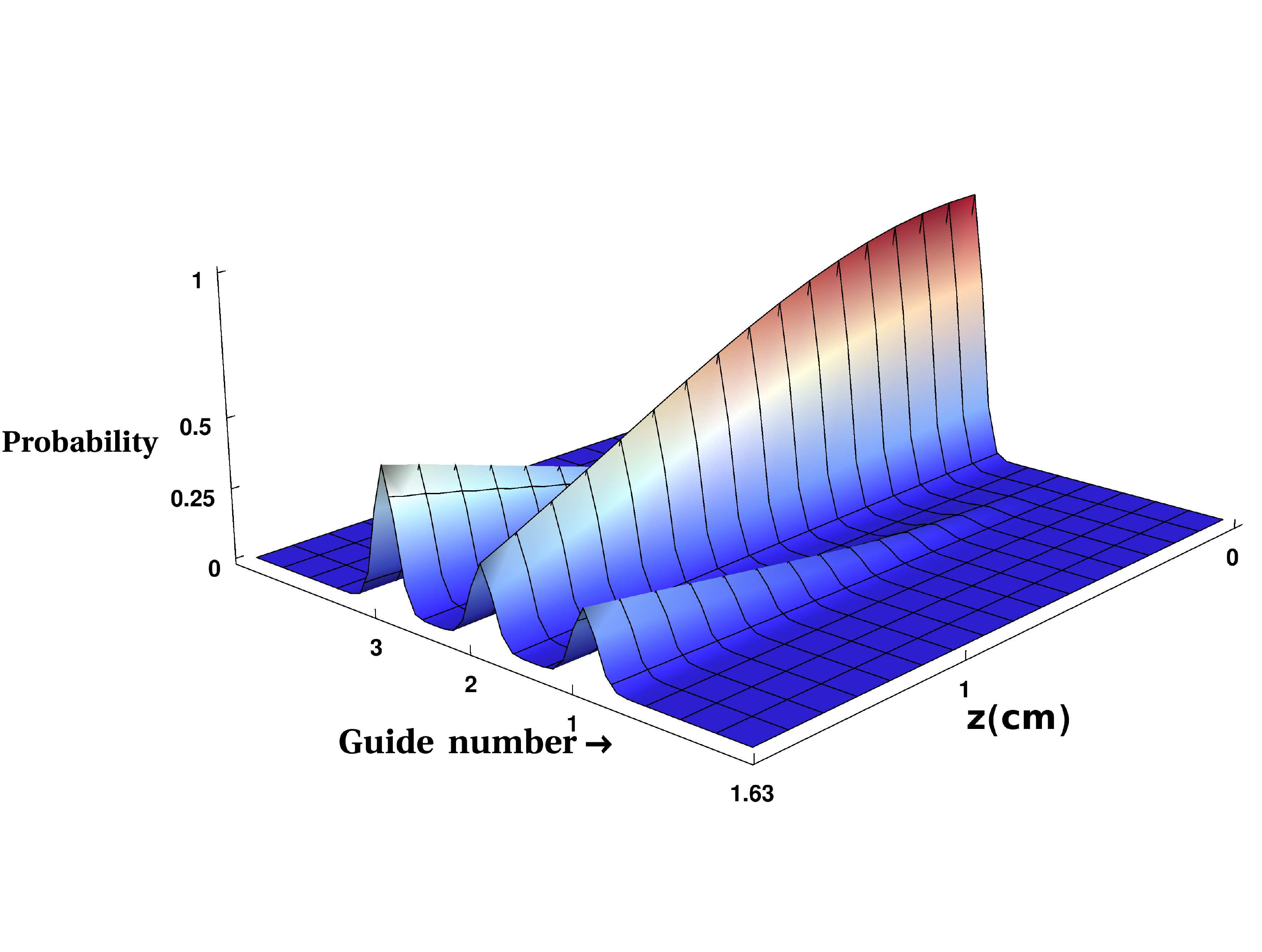}
  \caption{Evolution of wave function up to a distance of 1.634cm. At z=0 probability of finding the photon at middle guide was maximum(=1) and at z=1.634cm the probability becomes $\frac{1}{4}$, $\frac{1}{4}$, and $\frac{1}{2}$ at guides 1, 2, and 3 respectively.}
  \label{pl0}

\end{figure}
\begin{figure}[h]
  \includegraphics[width=0.47\textwidth]{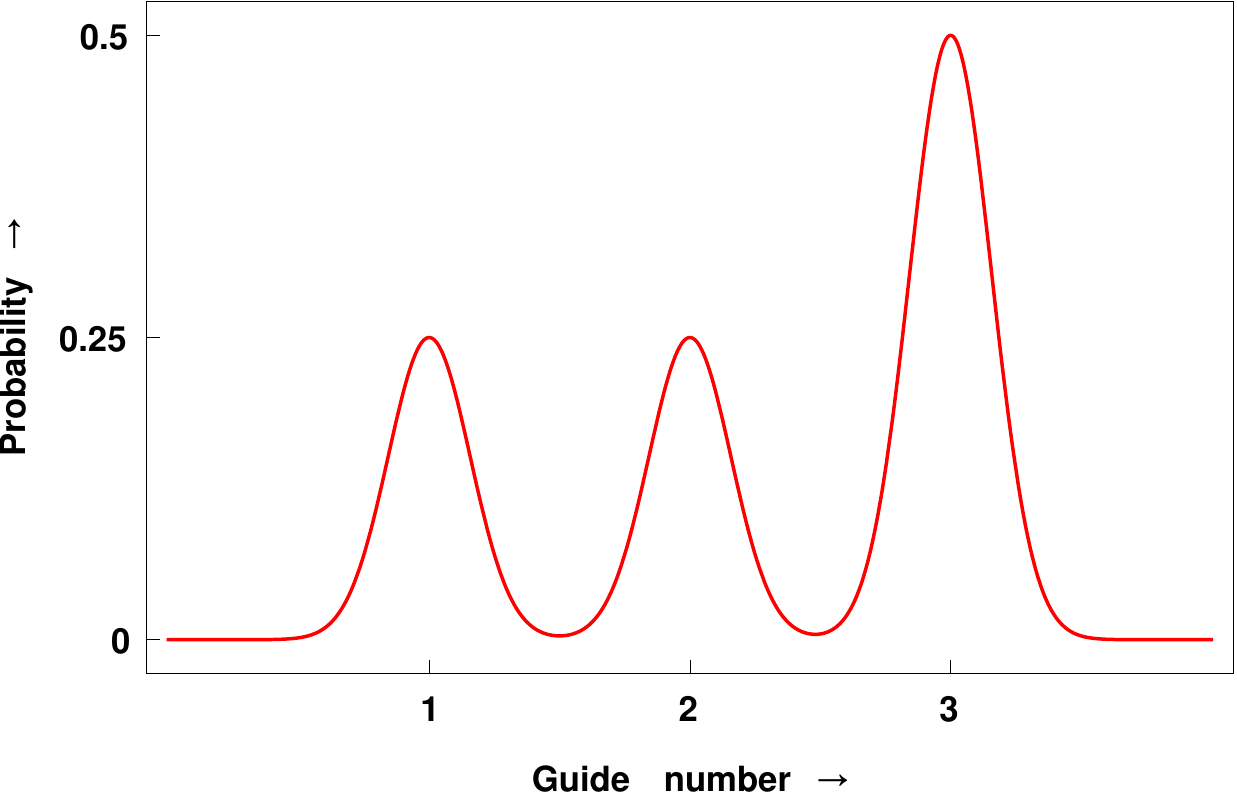}
  \caption{Probability at different guides at z=1.634cm.}
  \label{pl}
  \end{figure}
Now if a single photon is injected to the middle waveguide then after propagating a particular distance the photonic wave function will evolve to an asymmetric superposition and the required state will be obtained. The initial wave function of the system can be written as, $|\psi(0)\rangle=a_2^\dagger|000\rangle$ and it evolves to $|W_{p,1}\rangle=C_1|100\rangle$ + $C_2|010\rangle$ + $C_3|001\rangle$, where $C_1$, $C_2$ and $C_3$ are the probability amplitudes of $|100\rangle$, $|010\rangle$ and $|001\rangle$ respectively. The probability amplitudes are given as,
\begin{equation}\label{cp1}
 C_1=\frac{i \text{sin}[\sqrt{3}kz]}{\sqrt{3}}
\end{equation}
\begin{equation}\label{cp2}
C_2= \text{cos}[\sqrt{3}kz]
\end{equation}
\begin{equation}\label{cp3}
C_3=\frac{i \sqrt{2}}{\sqrt{3}}\text{sin}[\sqrt{3}kz]
\end{equation}

To obtain the desired state the parameters z and $k$ are to be selected in such a way that $|C_1|^2=\frac{1}{4}$, $|C_2|^2=\frac{1}{4}$ and $|C_3|^2=\frac{1}{2}$. The simultaneous solution of the above conditions yields $k$z=0.6046. This results in a state,
\begin{equation}\label{eq13}
 \arrowvert W_{p,1}\rangle= \frac{1}{2}(i\arrowvert100\rangle+ \arrowvert010\rangle+ i \sqrt{2}\arrowvert001\rangle)
\end{equation}

 The state represented in Eq. $\eqref{eq13}$ is obtained with a phase factor. Phase of such state in different modes can be manipulated inside the integrated system itself by addition of optical phase shifters~\cite{pnf_31} to individual guides. In particular, for our case a phase shifter can be added to the middle waveguide mode and an extra phase of $`i$' can be introduced. The resulting state will then be a perfect W-state with a global phase $`i$' i.e.,
\begin{equation}
  \arrowvert W_{p,1}\rangle= \frac{i}{2}(\arrowvert100\rangle+ \arrowvert010\rangle+  \sqrt{2}\arrowvert001\rangle)
\end{equation}

\begin{figure}[h]
  \includegraphics[width=0.5\textwidth]{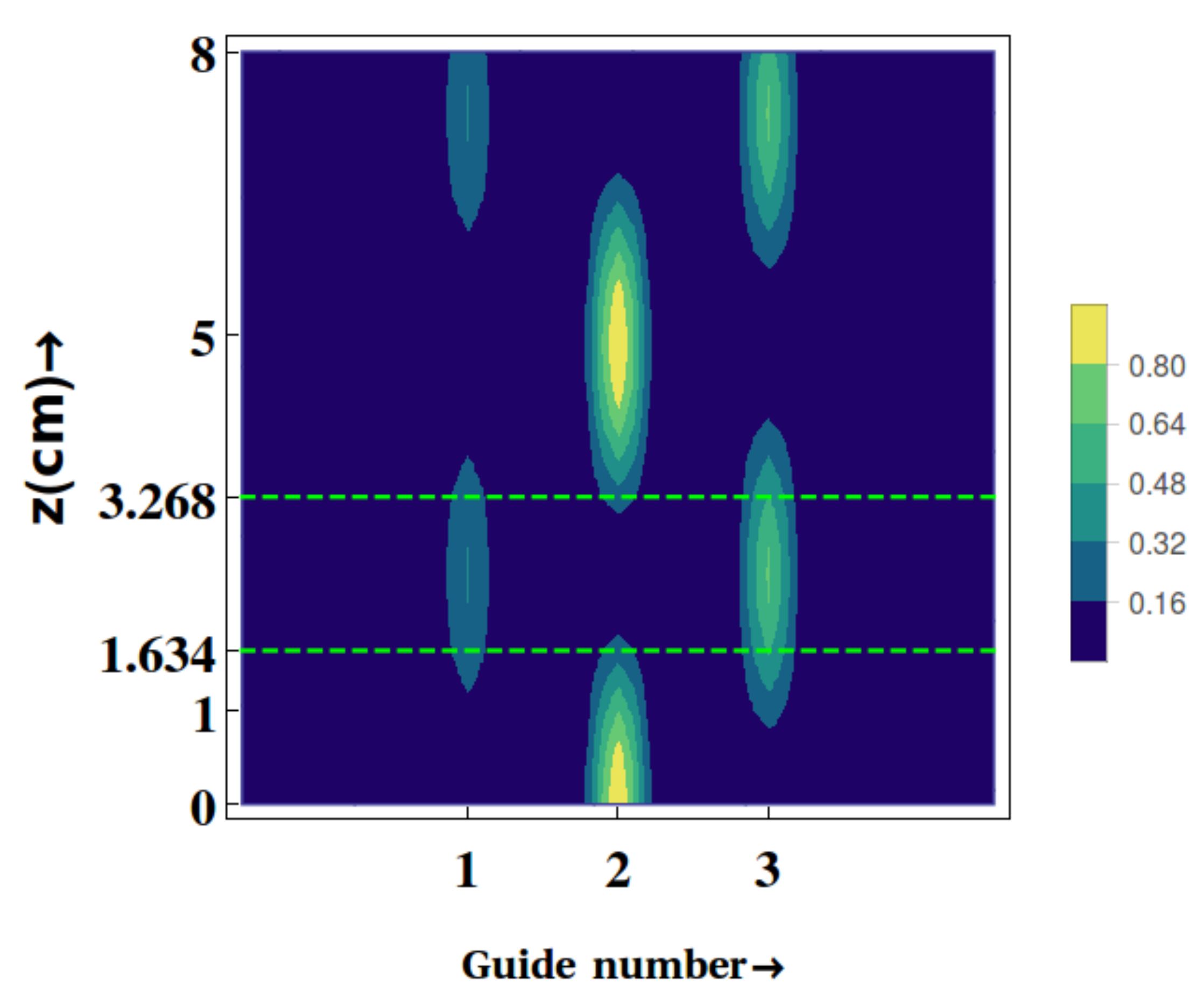}
  \caption{Contour plot showing probability in different guides with respect to propagation distance with dashed lines showing positions where perfect W-state is obtained.}
  \label{con}
\end{figure}

The integrated system of waveguides required to generate such a state is to be fabricated with relatively different coupling constant between adjacent guides but with same propagation constant. The coupling between two guides can be controlled by varying the distance between them. The dependence of coupling constant on distance for weakly coupled guides is given as $k_j'=
k e^{[-(d_j-d_1)/ d_0]}$ where $d_0$ and $d_1$ are fit parameters and $k$ is the characteristic coupling strength and the desired separation between the guides is given as, $d_j=d_1-d_0 \log (\sqrt{j})$. For example to obtain a $W_{p,1}$-state, coupling parameter $\gamma_j$ is chosen to be $\sqrt{j}$.
Hence the coupling coefficient between the guides become $k'_j=k\sqrt{j}$ with relative distance between the guides $d_j=d_1-d_0 \log (\sqrt{j})$. Such a coupling (as shown in Fig. \ref{wav}(b)) has already been realized using femtosecond laser direct writing method which results in stable and low loss waveguide structures known as Glauber-Fock photonic lattices~\cite{pnf_37}. The value of $k$z (normalized distance) is calculated to be 0.6046. For the type of system we have considered, the value of $k$ is given to be 0.37cm$^{-1}$ ~\cite{pnf_37}. Hence at a distance of 1.634cm, 3.268cm and so on, the initial wave function of the system evolves to perfect W-state after a phase modulation. The evolution of wave function to W-state at 1.634cm, is shown in Fig. \ref{pl0} and probability of each guide at 1.634cm is shown in Fig. \ref{pl}. The formation of W-state at different positions is shown by a contour plot in Fig. \ref{con}. The dashed lines represent the positions where the state is obtained.\\
\\

For s=2, $\Phi_1$=0 and $\Phi_2$=0 the form of perfect W-state is given as,
\begin{equation}
   \arrowvert W_{p,2}\rangle= \frac{1}{\sqrt{6}}(\arrowvert100\rangle+ \sqrt{2}\arrowvert010\rangle+ \sqrt{3}\arrowvert001\rangle)
\end{equation}
To obtain the state represented in the above equation, similar process can be followed. The parameter $\gamma_j$ for this state is chosen to be $\sqrt{2j-1}$ and the Heisenberg equation is given as,

\begin{equation}
 i \frac{d\hat{a}_j^\dagger}{dz}= k\sqrt{2j-3} \hat{a}_{j-1}^\dagger+ k \sqrt{2j-1}\hat{a}_{j+1}^\dagger
\end{equation}
and in matrix form,
\begin{equation}
i\frac{d}{dz}
 \begin{pmatrix}
\vspace{0.2cm} \hat{a}_1^\dagger \\ \vspace{0.2cm} \hat{a}_2^\dagger\\\hat{a}_3^\dagger

 \end{pmatrix}
= 
\begin{pmatrix}
 0 & k & 0\\k & 0 & \sqrt{3} k\\0 & \sqrt{3} k & 0
\end{pmatrix}
\begin{pmatrix}
 \vspace{0.2cm} \hat{a}_1^\dagger \\ \vspace{0.2cm} \hat{a}_2^\dagger\\\hat{a}_3^\dagger

\end{pmatrix}
\end{equation}

The evolution matrix is given as,
\begin{equation}
\scriptstyle e^{-izM}=
\begin{pmatrix}
\scriptscriptstyle\frac{1}{4}(3+\text{cos}[2kz]) & \scriptscriptstyle -i \text{cos}[kz] \text{sin}[kz]& \scriptscriptstyle -\frac{\sqrt{3}}{2}\text{sin}^{\tiny{2}}[kz]\\  \scriptscriptstyle -i \text{cos}[kz] \text{sin}[kz] &\scriptscriptstyle \text{cos}[2 kz] &  \scriptscriptstyle  -i \sqrt{3}  \text{cos}[kz] \text{sin}[kz]\\ \scriptscriptstyle-\frac{\sqrt{3}}{2}\text{sin}^2[kz] &  \scriptscriptstyle  -i\sqrt{3}  \text{cos}[kz] \text{sin}[kz] & \scriptscriptstyle \frac{1}{4}(1+3 \text{cos}[2 kz])
\end{pmatrix}
\end{equation}

Further proceeding in the same way as done for s=1, the value of $k$z where the required state is obtained, is calculated to be 0.4777. With the value of $k$=0.37cm$^{-1}$ W-state is obtained at 1.291cm. The coupling coefficient in this case can be given as, $k'_j=k\sqrt{2j-1}$. To achieve such coupling, the distance between the guides would become $d_j=d_1-d_0 \log (\sqrt{2j-1})$. Similarly the value of $\gamma_j$ for s=3 is chosen to be $\sqrt{3j-2}$. Hence the coupling coefficient and the distance between the guides can be given as  $k'_j=k\sqrt{3j-2}$ and $d_j=d_1-d_0 \log (\sqrt{3j-2})$ respectively.

Note that our scheme is based on the propagation of a single photon in coupled waveguide structure. The single photon required, is usually obtained by probabilistic non-linear spontaneous parametric down conversion (SPDC) process by the help of a non-linear crystal. Recently, single photons are obtained within the integrated system itself to avoid the use of bulk optical elements. Such type of deterministic single photon source has already been demonstrated in Ref.~\cite{pnf_38,pnf_39}.

\section{Discussion on state generation}
\subsection{Effect of dissipation}
In this section we consider the effect of photon loss. The generation scheme we have presented till now doesn't consider any loss. However a general analysis can be done by including the loss~\cite{oe1,oe2,oe3,oe4,oe5}.

\begin{figure}[h]
 \subfigure[]{\includegraphics[width=0.95\linewidth]{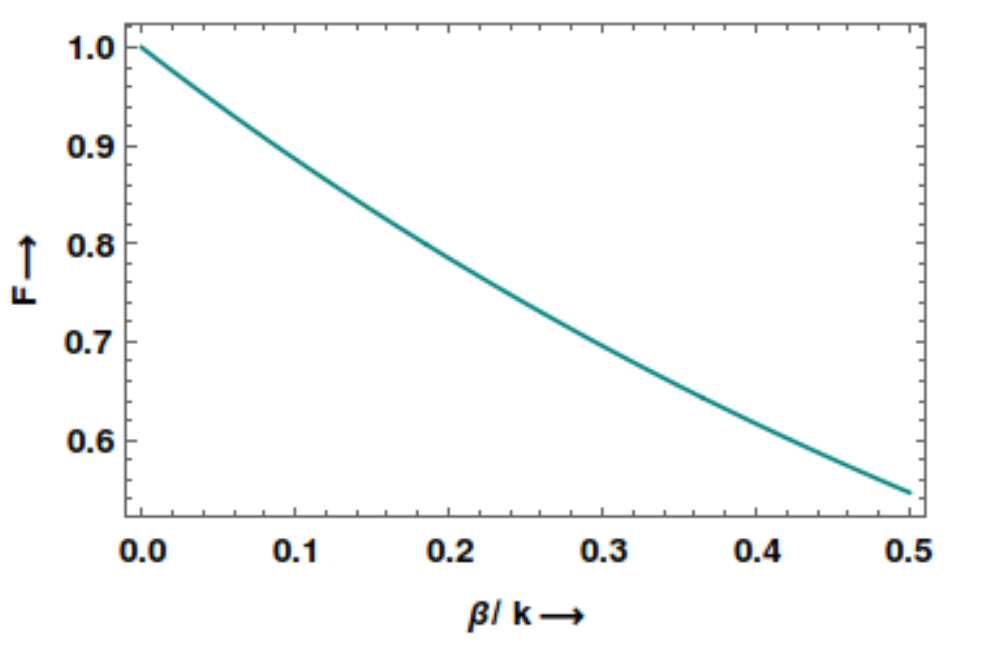}}
 \subfigure[]{\includegraphics[width=0.95\linewidth]{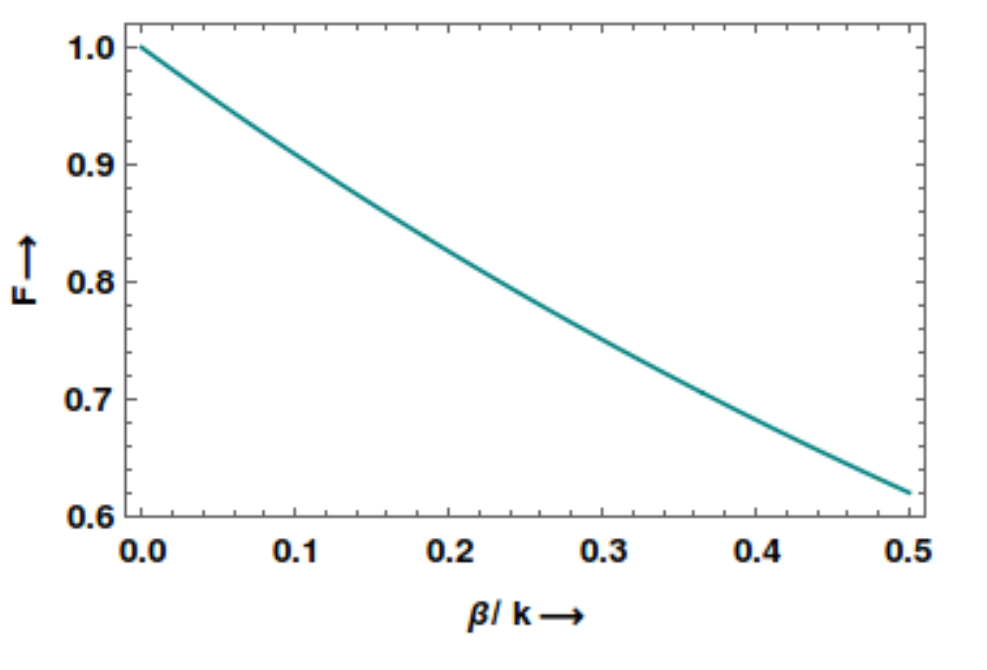}}
  
  \caption{Variation of fidelity with photon loss rate for (a) s=1 and (b) s=2 perfect W-state.}
  \label{fidelity}
\end{figure}

We study the effect of dissipation by using master equation given as,
\begin{equation}\label{master}
  \dot{\rho}=-\frac{i}{\hbar}[H,\rho] + \mathcal{L}\rho
\end{equation}
where $\rho$ is the time dependent density matrix of the state and $\mathcal{L}$ is the Lindblad superoperator which is given as,
\begin{equation}
    \mathcal{L\rho} = -\beta\sum_{j=1}^3 (\hat{a}^\dagger_j\hat{a_j}\rho-2\hat{a_j}\rho \hat{a}^\dagger_j+\rho\hat{a}^\dagger_j\hat{a_j} )
\end{equation}

where $\beta$ is the photon loss rate associated with each mode. Eq.  $\eqref{master}$ is solved for three modes to get the time dependent density matrix with loss factor, as done in Ref.~\cite{oe}.
The fidelity of state generation is then calculated by using the formula,
\begin{equation}
    F=[Tr(\sqrt{\sqrt{\sigma}\rho\sqrt{\sigma})}]^2\end{equation}

where $\sigma$ and $\rho$ are the density matrices obtained without and with loss. The variation of fidelity with loss for both case is shown in Fig. \ref{fidelity}. The typical values of $\beta/k$ are considered from  Ref.~\cite{oe}.
\\
From Fig. \ref{fidelity} it can be noted that the fidelity for state generation is not decreasing much with increase in $\beta/k$. Hence our scheme is robust in the presence of loss.
\subsection{Generalization of the scheme}
The generalized perfect W-state represented in Eq. $\eqref{eq2}$ contains various states. One can generalize the scheme for generation of any state with corresponding coupling and propagation distance. The generalized parameters can be obtained by comparing the generalized form of Eqs. $\eqref{cp1}$-$\eqref{cp3}$ with Eq. $\eqref{eq2}$. The parameters are given as,
\begin{equation}
    \gamma=\sqrt{s+1}
\end{equation}
and
\begin{equation}
    kz=\frac{1}{\sqrt{1+\gamma^2}}\tan^{-1}({\sqrt{\frac{1+\gamma^2}{s}})}
\end{equation}

where we set $\gamma$= $\gamma_2$ and $\gamma_1$=1. The parameters $\Phi_1$ and $\Phi_2$ appearing in Eq. $\eqref{eq2}$ can be manipulated inside the waveguides itself by the use of by optical phase control resistive elements~\cite{pnf_31}.

\section{Non-locality of perfect W-state}
In this section, the nonlocality of perfect W state (Eq. $\eqref{eq3}$) is demonstrated without and with inequality relation. The method based on Hardy-type proof of nonlocality without inequality, described for some generalized GHZ and W states~\cite{pnf_41}, is adopted here to prove the nonlocality of perfect W state without inequality relation. First, the observable corresponding to Pauli operator $\hat{Z}$ is considered. $\hat{Z}_i$ measurement at site $i$ with outcome $z_i = +1 (-1)$ indicates the presence (absence) of photon at that site. Since photon is present at only one site, the outcome $z_i = +1$ can't be obtained for more than one site, that is, $p(z_i = z_j = +1) = 0$. Suppose $\hat{Z}$ measurement is performed at sites $1$ and $2$. Then 
\begin{equation}\label{H}
z_1 = +1\:\: \text{and}\:\: z_2 = +1\:\: \text{never occurs}. 
\end{equation} 
Next, another operator $\hat{K}$ with eigenvectors $\vert k \pm \rangle$ corresponding to eigenvalues $\pm 1$, is considered. The basis vectors $\vert 1 \rangle$ and $\vert 0 \rangle$ can be expressed in the bases $\vert k \pm \rangle$ as follows. 
\begin{equation}
\vert 1 \rangle = \text{cos} \dfrac{\alpha}{2} \vert k + \rangle + \text{sin} \dfrac{\alpha}{2} \vert k - \rangle,
\end{equation}
\begin{equation}
\vert 0 \rangle = \text{sin} \dfrac{\alpha}{2} \vert k + \rangle - \text{cos} \dfrac{\alpha}{2} \vert k - \rangle.
\end{equation}
The states $\vert k \pm \rangle$ can be prepared similar to $\vert x \pm \rangle$ as described in Ref~\cite{heany}. In the above transformation, if the parameter, $\alpha$, is chosen to satisfy the following condition,
\begin{equation}
\text{cos}^2 \dfrac{\alpha}{2} = \sqrt{2} \text{sin}^2 \dfrac{\alpha}{2},
\end{equation}
the perfect W state can be rewritten in the bases $\{\vert k \pm \rangle\}_{2,3}$ and $\{\vert k \pm \rangle \}_{1,3}$ as follows:
\begin{equation}
\begin{aligned}
\vert W_{p,1} \rangle = \dfrac{1}{2} \text{sin}^2 \dfrac{\alpha}{2} \bigg\{ \vert 1 \rangle_i \bigg[ \vert k+ k+ \rangle + \sqrt{2} \vert k- k- \rangle \\ - \sqrt[4]{2} \bigg(\vert k+ k- \rangle + \vert k- k+ \rangle \bigg) \bigg] \\
 + \vert 0 \rangle_i \bigg[ \big( \sqrt[4]{2} + \sqrt[4]{8}\big) \bigg(\vert k+ k+ \rangle - \vert k- k- \rangle \bigg)\\ - \vert k- k+ \rangle \bigg] \bigg\}_{j,3},
 \end{aligned}
\end{equation}
where $i,j = 1,2$. 
In the basis $\{\vert k \pm \rangle\}_{2,3}$, $\hat{Z}$ measurement at site 1 and $\hat{K}$ measurement at sites 2 and 3 imply the following. 

\begin{equation}
\text{if}\:\:k_2 = +1,\:\: k_3 = -1 \:\: \text{then} \:z_1 = +1. 
\end{equation}

Similarly in the basis $\{\vert k \pm \rangle\}_{1,3}$, $\hat{Z}$ measurement at site 2 and $\hat{K}$ measurement at sites 1 and 3 imply 

\begin{equation}
\text{if}\:\:k_1 = +1,\:\: k_3 = -1 \:\: \text{then} \:z_2 = +1. 
\end{equation}

When the perfect W state is written in the basis $\{ \vert k \pm \rangle \}_{1,2,3}$ and $\hat{K}$ measurement is performed at all three sites, it can be verified that 

\begin{equation}
k_1 = k_2 = +1 \:\:\text{and}\:\:k_3 = -1 \:\:\text{sometimes occurs}. 
\end{equation}

\begin{figure}[h]
  \includegraphics[width=0.5\textwidth]{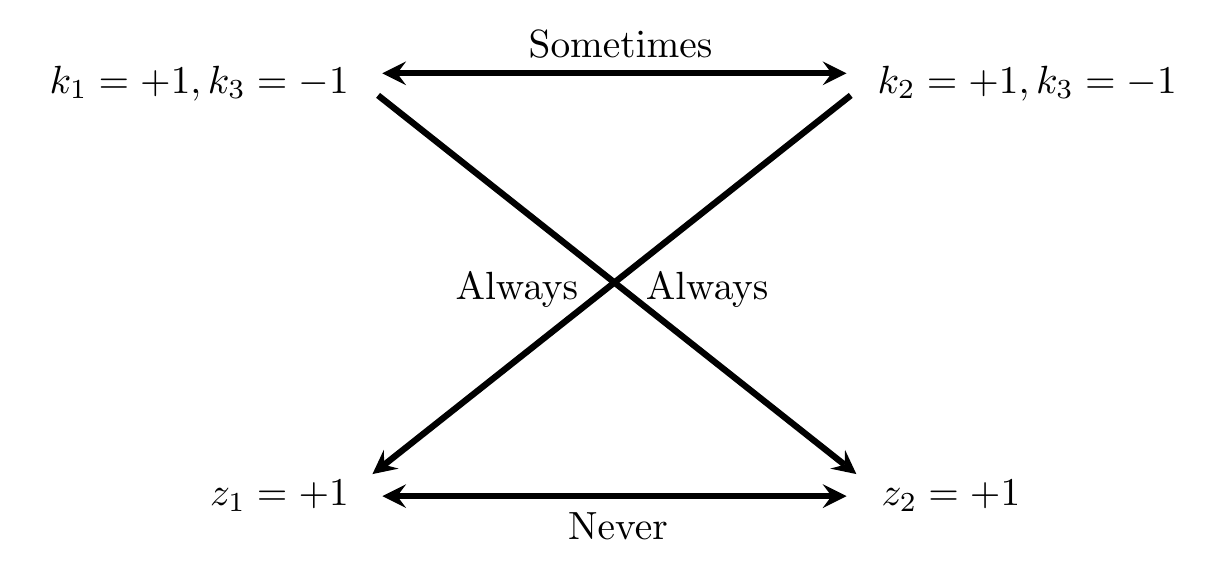}
  \caption{Diagram of proof of nonlocality without inequality for perfect W state.}
  \label{nonlocality}
\end{figure}

Let $\hat{K}$ measurement is performed at all three sites and the outcomes $k_1 = k_2 = +1$ and $k_3 = -1$ are obtained. In this particular measurement, if $\hat{Z}$ measurement had been performed at sites 1 and 2, instead of $\hat{K}$ measurement, by theory of local elements of reality, the outcomes at sites 1 and 2 would have been $z_1 = z_2 = +1$ which contradicts Eq. $\eqref{H}$. This clearly demonstrates the nonlocality of perfect W state.
This contradiction has "sometimes-always-never" structure (see Fig. \ref{nonlocality}) similar to Hardy's proof of nonlocality of nonmaximally entangled two qubit states~\cite{pnf_42,pnf_43}.

The probabilities involved in the above arguments can be written in the form of Bell-CH inequality~\cite{pnf_44,pnf_45} as follows.

\begin{equation}{\label{CH}}
\begin{aligned}
p(k_1 = +1, k_2=+1, k_3=-1) \\ - p(z_1=-1, k_2=+1, k_3=-1) \\
- p(k_1=+1, z_2=-1, k_3=-1) \\ - p(z_1 = +1, z_2=+1, k_3=-1) \leq 0.
\end{aligned}
\end{equation}

Violation of this inequality relation implies the nonlocality of the state. For perfect W state, the first probability in the above inequality relation can be found as 
\begin{equation}
p(k_1=+1, k_2=+1, k_3=-1) = \dfrac{1}{2} \text{sin}^6 \dfrac{\alpha}{2},
\end{equation}

and all other three probabilities are zero. Hence the perfect W state violates the inequality relation [Eq. $\eqref{CH}$] and it proves the nonlocal nature of the perfect W state.

\section{Conclusion}
In conclusion, we have proposed a method for generation of single photon three mode perfect W-state in a compact and interferometrically stable waveguide system, where the path degrees of freedom of a single photon has been manipulated to achieve the required state. The method we have proposed is experimentally feasible as the low loss factor ensures efficient generation of the state. Hence these systems can be regarded as a versatile tool to be used in quantum information processing. In future the system can be scaled accordingly to produce the generalized W-state~\cite{pnf_40} for quantum communication processes. We have also shown the nonlocality of single photon perfect W-state using local elements of reality.
\\
\vspace{2cm}\\


\begin{thebibliography}{50}


\bibitem{pnf_2}D.~M. Greenberger,  M.~A. Horne, and  A. Zeilinger, \emph{Bell's theorem, Quantum Theory, and Conceptions of the Universe}, Kluwer Academics, Dordrecht, The Netherlands (1989).
\bibitem{pnfl_4}A. Biswas and G. S. Agarwal, \href{\doibase 10.1080/09500340408232477}{J. Mod. Opt. \textbf{51}, 1627 (2004)}.
\bibitem{pnfl_25}R. H. Dicke, \href{\doibase 10.1103/PhysRev.93.99}{Phys. Rev. \textbf{93}, 99 (1954).}
\bibitem{pnf_7}W. D\"{u}r, G. Vidal, and J. I. Cirac, \href{\doibase 10.1103/PhysRevA.62.062314}{Phys. Rev. A \textbf{62}, 062314 (2000)}.
\bibitem{pnfl_26}M. Hein, J. Eisert, and H. J. Briegel, \href{\doibase 10.1103/PhysRevA.69.062311}{Phys. Rev. A \textbf{69}, 062311 (2004).}

\bibitem{pnf_1} D. Bouwmeester, J.~W. Pan,  M. Daniell,  H. Weinfurter, and A. Zeilinger, \href{\doibase  	10.1103/PhysRevLett.82.1345}{Phys. Rev. Lett. \textbf{82}, 1345 (1999)}.


\bibitem{pnf_3} J. S. Bell, \href{\doibase 10.1103/PhysicsPhysiqueFizika.1.195}{Physics \textbf{1}, 195 (1964)}.

\bibitem{pnf_4} X.~R. Jin, X. Ji,  Y.~Q. Zhang,  S. Zhang,  S.~K. Hong, K.~H. Yeon, and C.~I. Um, \href{\doibase 10.1016/j.physleta.2006.01.035}{Phys. Lett. A \textbf{354}, 67 (2006)}.

\bibitem{pnf_5}B. K. Behera, A. Banerjee, and P.~K. Panigrahi, \href{\doibase 10.1007/s11128-017-1762-0}{Quantum Inf. Process. \textbf{16}, 312 (2017).}

\bibitem{pnf_6}J.~C. Hao,  C.~F. Li, and G.~C. Guo, \href{\doibase 10.1103/PhysRevA.63.054301}{Phys. Rev. A \textbf{63}, 054301 (2001)}.





\bibitem{pnf_8} W. Jian, Z. Quan and T. Chao-Jing , \href{\doibase 10.1088/0253-6102/48/4/013}{Commun. Theor. Phys. \textbf{48} 637 (2007)}.
\bibitem{pnf_9} T. Hwang, C. C. Hwang, C. W. Tsai, \href{\doibase 10.1140/epjd/e2010-10320-y}{Eur. Phys. J. D \textbf{61}, 785 (2011)}.

\bibitem{pnfl_10}D. Gottesman, T. Jennewein, and S. Croke, \href{\doibase 10.1103/PhysRevLett.109.070503}{
Phys. Rev. Lett. \textbf{109}, 070503 (2012).}

\bibitem{pnf_10}D. Bru\ss, D. P. DiVincenzo, A. Ekert, C. A. Fuchs, C. Macchiavello, and J. A. Smolin, \href{\doibase 10.1103/PhysRevA.57.2368}{Phys. Rev. A \textbf{57}, 2368 (1998)}.
\bibitem{pnfl_1}A. Rai and G. S. Agarwal, \href{\doibase 10.1103/PhysRevA.79.053849}{Phys. Rev. A \textbf{79}, 053849 (2009)}.

\bibitem{w1}A. Perez-Leija \emph{et al.}, \href{\doibase 10.1103/PhysRevA.87.013842}{Phys. Rev. A \textbf{87}, 013842 (2013).}

\bibitem{w2}M. Gr\"{a}fe \emph{et al}, \href{\doibase 10.1038/nphoton.2014.204}{ Nature Photon \textbf{8}, 791 (2014).}


\bibitem{pnf_11} J. Joo, Y. J. Park, S. Oh, and J. Kim, \href{\doibase 10.1088/1367-2630/5/1/136}{New J. Phys. \textbf{5}, 136 (2003)}.

\bibitem{pnf_12} P. Agrawal and A. K. Pati, \href{\doibase 10.1103/PhysRevA.74.062320}{Phys. Rev. A. \textbf{74}, 062320 (2006)}.

\bibitem{pnf_13}S. B. Zheng, \href{\doibase doi.org/10.1103/PhysRevA.74.054303}{Phys. Rev. A \textbf{74}, 054303 (2006)}.
\bibitem{pnf_14} Y.-y. Nie, Y.-h. Li, J.-c. Liu, and M.-h. Sang, \href{\doibase 10.1016/j.optcom.2010.10.084}{Opt. Commun. \textbf{284}, 1457 (2011)}.



\bibitem{pnf_16}L.Dong \emph{et al}, \href{\doibase  10.1103/PhysRevA.93.012308}{Phys. Rev. A \textbf{93}, 012308 (2016)}.

\bibitem{pnf_32}S. M. Tan , D. F. Walls, and M. J. Collett , \href{\doibase 10.1103/PhysRevLett.66.252}{Phys. Rev. Lett. \textbf{66} 252 (1991).}

\bibitem{pnf_33}B. Hessmo, P. Usachev, H. Heydari and G. Björk, \href{\doibase 10.1103/PhysRevLett.92.180401} {Phys. Rev. Lett. \textbf{92} 180401 (2004).}

\bibitem{new1} S. Papp, K. Choi, H. Deng, P. Lougovski, S. van Enk, and H. Kimble, \href{\doibase 10.1126/science.1172260}{Science \textbf{324}, 764 (2009)}.

\bibitem{new2}F. Monteiro, V. C. Vivoli, T. Guerreiro, A. Martin, J.-D. Bancal, H. Zbinden, R. T. Thew, and N. Sangouard, \href{\doibase 10.1103/PhysRevLett.114.170504}{Phys.
Rev. Lett. \textbf{114}, 170504 (2015).}



\bibitem{new10}G. Björk, A. Laghaout, and U. L. Andersen, \href{\doibase 10.1103/PhysRevA.85.022316}{Phys. Rev. A \textbf{85}, 022316 (2012).}


\bibitem{new3}E. Lombardi, F. Sciarrino, S. Popescu, and F. Martini,
\href{\doibase 10.1103/PhysRevLett.88.070402}{Phys. Rev. Lett. \textbf{88}, 070402 (2002).}


\bibitem{new4}F. Sciarrino, E. Lombardi, G. Milani, and F. Martini,
\href{\doibase 10.1103/PhysRevA.66.024309}{Phys. Rev. A \textbf{66}, 024309 (2002).}

\bibitem{new5}D. Salart, O. Landry, N. Sangouard, N. Gisin, H. Herrmann, B. Sanguinetti, C. Simon, W. Sohler, R. T. Thew, A. Thomas, and H. Zbiden, \href{\doibase 10.1103/PhysRevLett.104.180504} {Phys. Rev. Lett. \textbf{104}, 180504 (2010).}



\bibitem{new11} T. Guerreiro, F. Monteiro, A. Martin, J. B. Brask, T. Vértesi,
B. Korzh, M. Caloz, F. Bussières, V. B. Verma, A. E. Lita,
R. P. Mirin, S. W. Nam, F. Marsilli, M. D. Shaw, N. Gisin,
N. Brunner, H. Zbinden, and R. T. Thew, \href{\doibase 10.1103/PhysRevLett.117.070404} {Phys. Rev.
Lett. \textbf{117}, 070404 (2016).}


\bibitem{new12}N. Sangouard, C. Simon, H. de Riedmatten, and N. Gisin, \href{\doibase 10.1103/RevModPhys.83.33}{ Rev. Mod. Phys. \textbf{83}, 33 (2011).}




\bibitem{r1}R. J. C. Spreeuw, \href{\doibase 10.1023/A:1018703709245}{Found. Phys. \textbf{28}, 361 (1998).}



\bibitem{r2} A. Aiello \emph{et al.}, \href{\doibase 10.1088/1367-2630/17/4/043024}{  New J. Phys. \textbf{17}, 043024 (2015).}

\bibitem{r3}S. Berg-Johansen \emph{et al.}, \href{\doibase 10.1364/OPTICA.2.000864}{Optica \textbf{2}, 864 (2015).}



\bibitem{r4} D. Guzman-Silva \emph{et al.}, \href{\doibase 10.1002/lpor.201500252}{Laser Photonics Rev. \textbf{10}, 317 (2016).}





\bibitem{pnf_17}A. Szameit and S. Nolte, \href{\doibase 10.1088/0953-4075/43/16/163001} {J. Phys. B: Mol. Opt,
Phys. \textbf{43} 163001 (2010)}.
\bibitem{pnf_18}T. Meany, M. Gr\"{a}fe , R. Heilmann, A. Perez-Leija , S. Gross , M.J. Steel,  M. J. Withford
A. Szameit, \href{\doibase 10.1002/lpor.201500061}{Laser \& Photonics Rev; \textbf{9}, 363 (2015)}.

\bibitem{pnf_19}M. Gr\"{a}fe \emph{et al}, \href{\doibase 10.1088/2040-8978/18/10/103002}{J. Opt. \textbf{18} 103002 (2016).}

\bibitem{pnf_20}J. L. O'Brien , A. Furusawa and  J. Vuckovic, \href{\doibase 10.1038/nphoton.2009.229}{Nat. Photon \textbf{3}, 687 (2009)}.

\bibitem{pnf_21}A. Politi,  M. J. Cryan,  J. G. Rarity,  S. Yu, and J. L. O'Brien, \href{\doibase 10.1126/science.1155441 }{Science \textbf{320}, 646 (2008)}.
\bibitem{pnf_22} A. Politi, J. C. F. Matthews, M. G. Thompson, and  J. L. O'Brien, \href{\doibase 10.1109/JSTQE.2009.2026060}{IEEE J. Sel. Top. Quantum Electron. \textbf{15}, 1673 (2009).}
\bibitem{pnf_23} L. Sansoni \emph{et al}, \href{\doibase 10.1103/PhysRevLett.105.200503}{Phys. Rev. Lett. \textbf{105}, 200503 (2010)}.
\bibitem{pnf_24} A. Crespi \emph{et al}, \href{\doibase 10.1038/ncomms1570}{Nat.Commun. \textbf{2}, 566 (2011).}
\bibitem{pnf_25}J. B. Spring \emph{et al}, \href{\doibase 10.1126/science.1231692}{Science \textbf{339}, 798 (2013)}.
\bibitem{pnf_26}A. Crespi \emph{et al}, \href{\doibase 10.1038/nphoton.2013.112}{Nat. Photon. \textbf{7}, 545 (2013).}
\bibitem{pnf_27}M. A. Broome \emph{et al}, \href{\doibase 10.1126/science.1231440 }{Science \textbf{339}, 794 (2013).}
\bibitem{pnf_28} M. Tillmann,\emph{et al}, \href{\doibase 10.1038/nphoton.2013.102}{Nat. Photon. \textbf{7}, 540 (2013).}
\bibitem{pnf_29}B. J. Metcalf,\emph{et al}, \href{\doibase 10.1038/nphoton.2014.217}{Nat. Photon. \textbf{8}, 770 (2014).}
\bibitem{pnf_30}A. Peruzzo, M. Lobino, J. C. F. Matthews, N. Matsuda, A. Politi, K. Poulios,  X. Zhou, Y. Lahini, N. Ismail, K. W\"{o}rhoff, Y. Bromberg, Y. Silberberg, M. G. Thompson and J. L. O'Brien, \href{\doibase 10.1126/science.1193515}{Science \textbf{329}, 1500 (2010).}

\bibitem{pnf_31}J. C. F. Matthews, A. Politi, A. Stefanov, and J. L. O'Brien, \href{\doibase 10.1038/nphoton.2009.93}{Nature Photonics \textbf{3}, 346 (2009).}






\bibitem{pnf_37}Robert keil \emph{et al}, \href{\doibase 10.1103/PhysRevLett.107.103601}{PRL \textbf{107}, 103601 (2011).}



\bibitem{pnf_38}Je-Hyung Kim \emph{et al}, \href{\doibase 10.1021/acs.nanolett.7b03220}{Nano Lett., \textbf{17} (12),  7394 (2017).}

\bibitem{pnf_39} P. Vergyris, T. Meany, T. Lunghi, G. Sauder, J. Downes,
M. J. Steel, M. J. Withford, O. Alibart, and S. Tanzilli, \href{\doibase 10.1038/srep35975}{Sci. Rep. \textbf{6}, 35975 (2016).}

\bibitem{oe}A. Rai, S. Das, and G. S. Agarwal, \href{\doibase 10.1364/OE.18.006241}{Opt. Express \textbf{18}, 6241
(2010).}

\bibitem{oe1}Z. Chen, Y. Zhou, and J.-T. Shen, \href{\doibase 10.1364/OL.42.000887}{Opt. Lett. 42, 887 (2017).}

\bibitem{oe2}Z. Chen, Y. Zhou, and J.-T. Shen, \href{\doibase 10.1103/PhysRevA.98.053830}{Phys. Rev. A \textbf{98}, 053830 (2018).}

\bibitem{oe3}Y. Shen, Z. Chen, Y. He, Z. Li, and J.-T. Shen,\href{\doibase 10.1364/JOSAB.35.000607}{J. Opt. Soc. Am. B \textbf{35}, 607 (2018).}

\bibitem{oe4}Z. Chen, Y. Zhou, and J.-T. Shen, \href{\doibase 10.1364/OL.41.003313} {Opt. Lett. \textbf{41}, 3313 (2016).}

\bibitem{oe5}Z. Chen, Y. Zhou, and J.-T. Shen, \href{\doibase 10.1103/PhysRevA.96.053805}{Phys. Rev. A \textbf{96}, 053805 (2017).}







\bibitem{pnf_41} R. Chang-Liang, S. Ming-Jun and D. Jiang-Feng \href{https://doi.org/10.1088/0256-307X/24/11/006}{Chin. Phys. Lett. \textbf{24}, 3036 (2007).}






\bibitem{heany}L. Heaney, A. Cabello, M. F. Santos, and V. Vedral, \href{\doibase 10.1088/1367-2630/13/5/053054}{New J. Phys. \textbf{13}, 053054 (2011).}










\bibitem{pnf_42}L. Hardy, \href{https://doi.org/10.1103/PhysRevLett.71.1665}{Phys. Rev. Lett. \textbf{71}, 1665 (1993).}

\bibitem{pnf_43}A. Cabello, \href{https://doi.org/10.1103/PhysRevA.65.032108}{Phys. Rev. A \textbf{65}, 032108 (2002).}

\bibitem{pnf_44}N. D. Mermin, \href{ https://doi.org/10.1119/1.17733}{Am. J. Phys. \textbf{62}, 880 (1994).}

\bibitem{pnf_45}L. Hardy, \href{https://doi.org/10.1103/PhysRevLett.73.2279}{Phys. Rev. Lett. \textbf{73}, 2279 (1994).} 

\bibitem{pnf_40} B. Pradhan, P. Agrawal, and A. K. Pati, \href{https://arxiv.org/abs/0805.2651}{arXiv:0805.2651 (2008).}
\bibliography{\jobname}
\end{thebibliography}
\end{document}